\documentstyle[prd,aps,epsf,12pt]{revtex}
\begin{document} 

\tighten
\draft
\preprint{DAMTP96-112} 

\title{The Quantum Canonical Ensemble} 
 
\author{
Dorje C. Brody$^{*}$ 
and 
Lane P. Hughston$^{\dagger}$ 
} 
\address{$*$DAMTP, Silver Street, Cambridge CB3 9EW U.K. \\ 
and Churchill College, Cambridge CB3 0DS U.K.} 
\address{$\dagger$Merrill Lynch International, 
25 Ropemaker Street, London EC2Y 9LY U.K. \\ 
and King's College London, The Strand, London WC2R 2LS U.K.} 

\date{\today} 

\maketitle 

\begin{abstract} 
The phase space ${\sl \Gamma}$ of quantum mechanics can be viewed 
as the complex projective space $CP^{n}$ endowed with a K\"ahlerian 
structure given by the Fubini-Study metric and an associated 
symplectic form. We can then interpret the Schr\"odinger equation 
as generating a Hamiltonian dynamics on ${\sl \Gamma}$. Based upon 
the geometric structure of the quantum phase space we introduce 
the corresponding natural microcanonical and canonical ensembles. 
The resulting density matrix for the canonical 
${\sl \Gamma}$-ensemble differs from density matrix of the 
conventional approach. As an illustration, the results are applied 
to the case of a spin one-half particle in a heat bath with an 
applied magnetic field. 
\end{abstract} 



\section{Quantum phase space} 

With a view to certain geometrical generalisations we reformulate 
the conventional quantum theory in the following manner. Consider 
a finite dimensional complex Hilbert space ${\cal H}_{\bf C}^{n+1}$ 
with complex elements $Z^{\alpha}$. We identify the state vector 
$Z^{\alpha}$ with its multiples $\lambda Z^{\alpha}$ 
($\lambda\in{\bf C}-\{0\}$) to obtain the complex projective space 
$CP^{n}$. Here we use Greek indices ($\alpha=0,1,\cdots,n$) for 
vectors in ${\cal H}_{\bf C}^{n+1}$, and regard $CP^{n}$ as the 
true `state space' of quantum mechanics. The complex conjugate 
of $Z^{\alpha}$ is written ${\bar Z}_{\alpha}$, with the 
Hermitian inner product $Z^{\alpha}{\bar Z}_{\alpha}$. We 
let the state vector $Z^{\alpha}$ represent homogeneous 
coordinates for $CP^{n}$. Then, the complex conjugate of a point 
$P^{\alpha}$ in $CP^{n}$ corresponds to the hyperplane 
${\bar P}_{\alpha}Z^{\alpha} = 0$. The points on this plane 
are the states orthogonal to the original state $P^{\alpha}$. 
Thus, $CP^{n}$ is equipped with a complex conjugation operation 
that maps points to hyperplanes of codimension one, and vice-versa. 
\par 

Distinct points $X^{\alpha}$ and $Y^{\alpha}$ are joined by a 
complex projective line $L^{\alpha\beta} = X^{[\alpha}Y^{\beta]}$, 
representing complex superpositions of the original two states. 
The quantum mechanical transition probability between two states 
$X^{\alpha}$ and $Y^{\alpha}$ is the cross ratio $\kappa = 
X^{\alpha}{\bar Y}_{\alpha}Y^{\beta}{\bar X}_{\beta}/X^{\gamma}
{\bar X}_{\gamma}Y^{\delta}{\bar Y}_{\delta}$. If the system is in 
the state $Y^{\alpha}$ and a measurement is made with the projection 
operator $X^{\alpha}{\bar X}_{\beta}/X^{\gamma}{\bar X}_{\gamma}$, 
then $\kappa$ is the probability that the result is the eigenvalue 
unity. The angle $\theta$ defined by $\kappa=\cos^{2}(\theta/2)$ can 
be interpreted as the distance between the states $X^{\alpha}$ and 
$Y^{\alpha}$. Then if we set $\theta=ds$, $X^{\alpha} = Z^{\alpha}$, 
and $Y^{\alpha} = Z^{\alpha}+dZ^{\alpha}$, retaining terms to second 
order, we obtain the unitarily invariant Fubini-Study metric 
\cite{kibble,gibbons,ashtekar} on the state space, given by 
\begin{equation} 
ds^{2}\ =\ 8(Z^{\alpha}{\bar Z}_{\alpha})^{-2} 
Z^{[\alpha}dZ^{\beta]}{\bar Z}_{[\alpha}d{\bar Z}_{\beta]}\ .
\end{equation} 
Now suppose we write $H^{\alpha}_{\beta}$ for the Hamiltonian 
operator, assumed Hermitian. Then, if we set $\hbar=1$, the 
Schr\"odinger equation reads 
$dZ^{\alpha} = iH^{\alpha}_{\beta}Z^{\beta}dt$. 
However, the projective Schr\"odinger equation 
\begin{equation} 
Z^{[\alpha}dZ^{\beta]}\ =\ 
iZ^{[\alpha}H^{\beta]}_{\gamma}Z^{\gamma}dt\ , 
\end{equation} 
which eliminates the superfluous freedom associated with the scale 
of $Z^{\alpha}$, is defined directly on the state space. \par 

An alternative way of viewing this structure is to regard $CP^{n}$ 
as a real manifold ${\sl \Gamma}$ of dimension $2n$, equipped with 
a Riemannian metric $g_{ab}$ and a compatible symplectic structure 
$\Omega_{ab}$, satisfying $g^{ab}\Omega_{ac}\Omega_{bd} = g_{cd}$ 
and $\nabla_{a}\Omega_{bc}=0$. We use Roman indices (${a} = 1, 2, 
\cdots, 2n$) for tensorial operations in the tangent space of 
${\sl \Gamma}$. The Schr\"odinger equation then takes the form of 
a Hamiltonian flow $\xi^{a} = 2\Omega^{ab}\nabla_{b}H$ on ${\sl 
\Gamma}$. Here the generating function $H(x)$ is given at each 
point $x$ in ${\sl \Gamma}$ by the expectation $H^{\alpha}_{\beta}
\Pi^{\beta}_{\alpha}(x)$ of the Hamiltonian operator, where 
$\Pi^{\beta}_{\alpha}(x) = Z^{\beta}{\bar Z}_{\alpha}/
Z^{\gamma}{\bar Z}_{\gamma}$ is the projection operator for the point 
$x\in{\sl\Gamma}$ represented by the state vector $Z^{\alpha}(x)$. 
Thus, in quantum mechanics the dynamical trajectories are given by a 
symplectic flow on the quantum phase space ${\sl \Gamma}$, generated 
by the Hamiltonian function $H(x)$. The flow is necessarily a Killing 
field, satisfying 
\begin{equation} 
\nabla_{(a}\xi_{b)}\ =\ 0\ , 
\end{equation} 
where $\xi_{a} = g_{ab}\xi^{b}$. In other words, the isometries 
of the Fubini-Study metric on ${\sl\Gamma}$ can be lifted to 
${\cal H}^{n+1}_{\bf C}$ to yield unitary transformations. 
Conversely, given the Killing field we recover the observable 
function $H(x)$ on ${\sl \Gamma}$, up to an additive constant, by 
use of the relation 
\begin{equation} 
\Omega^{ab}\nabla_{a}\xi_{b}\ =\ 2(n+1)(H-{\bar H})\ , 
\end{equation} 
where ${\bar H} = H^{\alpha}_{\alpha}/(n+1)$ is the average of the 
eigenvalues of $H^{\alpha}_{\beta}$. \par 

\section{Quantum microcanonical postulates} 

In this paper we use the metrical symplectic geometry of the quantum 
phase space as the basis for a reexamination of the traditional 
hypotheses of quantum statistical mechanics. To this end, we observe 
that the quantum phase space ${\sl \Gamma}$ admits a natural 
foliation by the $(2n-1)$-dimensional hypersurfaces ${\cal E}_{E}$ 
determined by level values $H(x)=E$ of the Hamiltonian function. The 
number of quantum mechanical microscopic configurations (pure states) 
with expected energy in the small range $E$ to $E+\Delta E$ is then 
given by $\Omega(E)\Delta E$, where the state density $\Omega(E)$ for 
energy $E$ is 
\begin{equation} 
\Omega(E)\ =\ \int_{{\cal E}_{E}} 
\frac{\nabla_{a}H d\sigma^{a}}{(\nabla_{b}H\nabla^{b}H)}\ . 
\label{eq:sd} 
\end{equation} 
Here $d\sigma^{a}=g^{ab}\epsilon_{bc\cdots d}dx^{c}\cdots dx^{d}$ 
is the natural vector-valued ($2n-1$)-form on ${\sl \Gamma}$. 
In the case of an isolated quantum mechanical system with energy in 
the given small range, we can adopt the notion of the microcanonical 
ensemble, and identify the entropy by use of the Boltzmann relation 
$S(E) = \ln(\Omega(E)\Delta E)$. Here, we implicitly assume what 
might be called the {\it quantum microcanonical postulate}, which 
asserts that for an isolated system in equilibrium all states on a 
given energy surface in the quantum phase space are equally probable. 
As a consequence, the temperature of such a system is given by $\beta 
= dS(E)/dE$, where $\beta = 1/kT$. Thus for an isolated quantum 
system with expected energy $E$ the equilibrium configuration is 
given by a uniform distribution on the energy surface ${\cal E}_{E}$, 
with entropy $S(E)$ and temperature $T(E)$ as given above. The 
corresponding probability density function on ${\sl\Gamma}$, which 
we call the microcanonical ${\sl\Gamma}$-distribution, is 
\begin{equation} 
\mu_{E}(x)\ =\ \delta(H(x)-E)/\Omega(E)\ . 
\end{equation} 
It is a straightforward exercise to verify that the state density 
$\Omega(E) = \int_{\Gamma}\delta(H(x)-E)dV$, which appears here as 
a normalisation factor, is consistent with expression 
(\ref{eq:sd}). \par 

A general measurable function $F(x)$ on ${\sl\Gamma}$ represents 
a nonlinear observable in the sense of \cite{kibble,weinberg}. We 
interpret $F(x)$ as the conditional expectation $\langle 
F\rangle_{x}$ of the observable $F$ in the pure state $x$. The 
unconditional expectation of $F$ in the microcanonical 
${\sl\Gamma}$-ensemble is then given by $\langle F\rangle_{E} = 
\int_{\sl\Gamma}F(x)\mu_{E}(x)dV$. In the case of a conventional 
linear observable we have $F(x) = F_{\alpha}^{\beta}
\Pi_{\beta}^{\alpha}(x)$. Then the unconditional expectation in the 
state $\mu_{E}(x)$ is $\langle F\rangle_{E}=F_{\alpha}^{\beta}
\mu^{\alpha}_{\beta}(E)$, where the quantum microcanonical density 
matrix $\mu^{\alpha}_{\beta}(E)$, parameterised by $E$, is 
\begin{equation} 
\mu^{\alpha}_{\beta}(E)\ =\ \int_{\sl\Gamma} 
\Pi^{\alpha}_{\beta}(x)\mu_{E}(x)dV\ . 
\end{equation} 
Providing we only consider linear observables, the state 
is fully characterised by $\mu^{\alpha}_{\beta}(E)$. Now suppose 
$W(E)$ denotes the total phase space volume for states such that 
$H(x)\leq E$. Then the density matrix $\mu^{\alpha}_{\beta}(E)$ 
can be calculated explicitly by use of the formula 
\begin{equation} 
\mu^{\alpha}_{\beta}(E)\ =\ 
\left(\frac{\partial W(E)}{\partial{\bar H}}\right)^{-1} 
\frac{\partial W(E)}{\partial H^{\beta}_{\alpha}}\ . 
\end{equation} 
\par 

\section{Canonical $\Gamma$-ensemble} 

Next, we motivate a construction for the corresponding canonical 
ensemble, representing the situation where a quantum mechanical 
system is in contact with a heat bath at a fixed temperature. The 
results we obtain are related to those that follow from the 
conventional density matrix for the canonical state, though differ 
in certain essential 
respects. First we note that the projection operator 
$\Pi^{\alpha}_{\beta}(y)$, for a given state $y$, can itself be 
thought of as the density matrix corresponding to a mass 
distribution $\delta_{y}(x)$ on ${\sl\Gamma}$, concentrated at the 
point $y$. Therefore, we can write $\Pi^{\alpha}_{\beta}(y) = 
\int_{\sl\Gamma}\Pi^{\alpha}_{\beta}(x)\delta_{y}(x)dV$. Then the 
conventional density matrix of quantum statistical mechanics, 
which we denote $r_{\beta}^{\alpha}$, is determined by a point mass 
distribution with density 
\begin{equation} 
r(x)\ =\ Z(\beta)^{-1} \sum_{k}e^{-\beta E_{k}}
\delta_{y_{k}}(x) 
\end{equation} 
on ${\sl\Gamma}$, concentrated at the energy eigenstates $y_{k}$, 
with $E_{k}$ being the corresponding eigenvalues and $Z(\beta) = 
\sum_{k}e^{-\beta E_{k}}$ the associated partition function. Hence, 
for the conventional density matrix we have 
\begin{equation} 
r_{\beta}^{\alpha}\ =\ Z(\beta)^{-1} \sum_{k} 
e^{-\beta E_{k}}\Pi^{\alpha}_{\beta}(y_{k})\ . 
\end{equation} 
It should be evident that, despite its general usefulness and wide 
acceptance as a fundamental basis for quantum statistical mechanics, 
the conventional density matrix $r^{\alpha}_{\beta}$ is in some 
respects unnatural in the absence of some mechanism forcing the 
system to energy eigenstates, since its phase space distribution is 
characterised by $\delta_{y_{k}}(x)$. Thus we are led to consider 
an alternative expression, suggested by the geometry of 
${\sl\Gamma}$, in which we do not assume a concentration on 
eigenstates. \par 

Consider two systems I and II and let them make contact to form a 
combined system I+II. We suppose that the systems only interact 
weakly, so the support of the resulting state on the quantum phase 
space ${\sl \Gamma}_{1+2}$ of the combined system is concentrated 
on the product ${\sl\Gamma}_{1\cdot2}\subset{\sl\Gamma}_{1+2}$ of 
the phase spaces 
${\sl \Gamma}_{1}$ and ${\sl \Gamma}_{2}$ for the systems I and II. 
We have in mind the situation where system I represents a heat bath 
with a given inverse temperature $\beta$, while II represents a small 
system immersed in the bath. We wish to calculate the probability 
density $p(E_{2})$ that system II lies on the energy surface 
${\cal E}_{E_{2}}$ in ${\sl \Gamma}_{2}$, conditional on the total 
energy of the combined system I+II lying in the small range $E$ to 
$E+\Delta E$. Now, according to the quantum microcanonical 
hypothesis, the equilibrium distribution over the energy surface 
${\cal E}_{E}$ in ${\sl \Gamma}_{1\cdot2}$ is uniform. Therefore, 
conditional on a given value $E$ for the combined system I+II, the 
joint probability density function $p_{E}(E_{1},E_{2})$ for system 
I to lie on ${\cal E}_{E_{1}}$ in ${\sl \Gamma}_{1}$ and for system 
II to lie on ${\cal E}_{E_{2}}$ in ${\sl \Gamma}_{2}$ is 
\begin{equation} 
p_{E}(E_{1},E_{2})\ =\ \frac{\delta(E-E_{1}-E_{2})
\Omega_{1}(E_{1})\Omega_{2}(E_{2})}{\Omega_{1\cdot2}(E)}\ . 
\end{equation} 
Here, $\Omega_{1}(E_{1})$ and $\Omega_{2}(E_{2})$ are the state 
densities for the energy surfaces ${\cal E}_{E_{1}}$ in 
${\sl \Gamma}_{1}$ and ${\cal E}_{E_{2}}$ in ${\sl \Gamma}_{2}$, and the 
state density $\Omega_{1\cdot2}(E)$ for the energy surface 
${\cal E}_{E}$ in ${\sl \Gamma}_{1\cdot2}$ is 
\begin{equation} 
\Omega_{1\cdot2}(E)\ =\ \int_{-\infty}^{\infty}\Omega_{1}(E-\epsilon)
\Omega_{2}(\epsilon)d\epsilon \ . 
\end{equation} 
Our goal is to calculate the conditional probability density for 
system II to have energy $E_{2}$. This is given by $p(E_{2}) = 
\int_{-\infty}^{\infty}p_{E}(E_{1},E_{2})dE_{1}$, which implies 
\begin{equation} 
p(E_{2})\ =\ \frac{\Omega_{1}(E-E_{2})\Omega_{2}(E_{2})}
{\Omega_{1\cdot2}(E)}\ . 
\end{equation} 
Since system I, the heat bath, is in equilibrium, by 
the microcanonical hypothesis it has entropy $S_{1}(E_{1}) = 
\ln(\Omega_{1}(E_{1})\Delta E)$, and hence temperature $\beta(E_{1}) 
= \partial S_{1}/\partial E_{1}$. If $E_{2}\ll E$, then to first 
order in $E_{2}$ we have $\Omega_{1}(E-E_{2}) = 
\Omega_{1}(E)e^{-\beta E_{2}}$, where $\beta(E_{1})$ is the inverse 
temperature of the heat bath. As a consequence we deduce that the 
probability density $p(E_{2})$ for system II to lie on the energy 
surface ${\cal E}_{E_{2}}$ in ${\sl \Gamma}_{2}$ is given by 
$p(E_{2}) = \Omega_{2}(E_{2})e^{-\beta E_{2}}/\int_{-\infty}^{\infty}
\Omega_{2}(\epsilon)e^{-\beta\epsilon} d\epsilon$. More succinctly, 
now we drop the subscript $2$, and regard the conditioning as 
implicit in the specification of the parameter $\beta$. Then for the 
probability density of the energy distribution on ${\sl\Gamma}$ 
we have 
\begin{equation} 
p(\epsilon)\ =\ \frac{\Omega(\epsilon)
\exp(-\beta\epsilon)}{Z(\beta)}\ , 
\label{eq:cano3} 
\end{equation} 
where $\Omega(\epsilon)$ is the state density on ${\sl\Gamma}$ per 
unit of energy and $Z(\beta)=\int_{-\infty}^{\infty}\Omega(\epsilon)
\exp(-\beta\epsilon)d\epsilon$ is the partition function. By the 
{\it quantum canonical postulate} we mean the assumption that a 
small quantum system in equilibrium with a heat bath will be in a 
state characterised by a distribution over the energy surfaces of 
${\sl\Gamma}$ with density (\ref{eq:cano3}), having a uniform 
distribution on each such surface. The corresponding probability 
density function $\rho_{\beta}(x)$ on ${\sl \Gamma}$ is 
\begin{equation} 
\rho_{\beta}(x)\ =\ \frac{\exp(-\beta H(x))}{Z(\beta)}\ , 
\label{eq:cano4} 
\end{equation} 
which we call the {\it canonical ${\sl\Gamma}$-distribution}. 
Alternatively, we can write $\rho_{\beta}(x) = 
\int_{\epsilon}p(\epsilon)\mu_{\epsilon}(x)d\epsilon$, expressing 
$\rho_{\beta}(x)$ as a weighted average of microcanonical ensembles, 
where $p(\epsilon)$ is given as in (\ref{eq:cano3}). The associated 
density matrix for the canonical ${\sl\Gamma}$-distribution is 
therefore 
\begin{equation} 
\rho^{\alpha}_{\beta}(T)\ =\ \int_{\sl\Gamma}
\Pi^{\alpha}_{\beta}(x)\rho_{\beta}(x)dV\ . 
\end{equation} 
\par 

\section{Two-state system} 

To illustrate how the consequences of the canonical 
${\sl\Gamma}$-distribution (\ref{eq:cano4}) differ from those of 
the conventional treatment in terms of the density matrix 
$r^{\alpha}_{\beta}$, we study the example of a spin one-half 
particle in a heat bath. First we review briefly the geometry of 
the Schr\"odinger dynamics for a spin one-half particle. The 
quantum phase space in this case is 
$CP^{1}$, which can be viewed as a sphere $S^{2}$. The poles of 
the sphere correspond to the energy eigenstates, 
with eigenvalues $E_{0}=-h$ and $E_{1}=h$, and the Schr\"odinger 
trajectories are generated by a rigid rotation of the sphere about 
the axis through the poles, the two stationary points. This rotation 
gives rise to a Killing field on the sphere, where the angular 
velocity is $2h$, the energy difference between the poles. Thus the 
Schr\"odinger evolution generates a latitudinal circle. Now we 
introduce a complex Hilbert space with coordinates $Z^{\alpha}$ 
$(\alpha = 0,1)$ which we regard as homogeneous coordinates for 
$CP^{1}$. The complex conjugate of $Z^{\alpha}$ is the `plane' 
${\bar Z}_{\alpha}$ in $CP^{1}$, which in this dimension is a point. 
The point corresponding to ${\bar Z}_{\alpha}$ is ${\bar Z}^{\alpha} 
= \epsilon^{\alpha\beta}{\bar Z}_{\beta}$, where 
$\epsilon^{\alpha\beta}$ is the natural symplectic form on the 
two-dimensional Hilbert space. The formalism in this case is 
equivalent to the algebra of two-component spinors. By use of 
the spinor identity $2X^{[\alpha}Y^{\beta]} = 
\epsilon^{\alpha\beta}X_{\gamma}Y^{\gamma}$, where 
$X^{\alpha}\epsilon_{\alpha\beta}=X_{\beta}$, we obtain 
$ds^{2} = 4Z_{\alpha}dZ^{\alpha}{\bar Z}_{\beta}d{\bar Z}^{\beta}/
({\bar Z}_{\gamma}Z^{\gamma})^{2}$ for the Fubini-Study metric, 
and $Z_{\gamma}dZ^{\gamma} = i H_{\alpha\beta}Z^{\alpha}Z^{\beta}dt$ 
for the projective Schr\"odinger equation. The Hamiltonian has the 
representation 
\begin{equation} 
H_{\alpha\beta}\ =\ 2h\frac{P_{(\alpha}{\bar P}_{\beta)}}
{{\bar P}_{\gamma}P^{\gamma}}\ , 
\end{equation} 
where $P^{\alpha}$ and ${\bar P}^{\alpha}$ are the stationary 
points, satisfying $H^{\alpha}_{\beta}P^{\beta}=hP^{\alpha}$ and 
$H^{\alpha}_{\beta}{\bar P}^{\beta}=-h{\bar P}^{\alpha}$. It follows 
that the velocity of the trajectory through ${\sl\Gamma}$ is 
\begin{equation} 
\frac{ds}{dt}\ =\ 2h\sin\theta\ , 
\end{equation} 
a special case of the Anandan-Aharonov relation \cite{aa}. Here, 
$\theta$ is 
the distance from $Z^{\alpha}$ to $P^{\alpha}$, given by the angular 
coordinate measured down from the north pole. The transition 
probability from $Z^{\alpha}$ to the north pole $P^{\alpha}$ is 
$(1 + \cos\theta)/2$, and for the evolutionary trajectory we obtain 
\begin{equation} 
 Z^{\alpha}\ =\ \cos(\theta/2)e^{i(ht+\phi)}P^{\alpha} 
+ \sin(\theta/2)e^{-i(ht+\phi)}{\bar P}^{\alpha}\ , 
\end{equation} 
where $\theta$ and $\phi$ are the initial coordinates on 
the sphere for the state at $t=0$. The vector $Z^{\alpha}$ 
is normalised by setting $P^{\alpha}{\bar P}_{\alpha}= 
-P_{\alpha}{\bar P}^{\alpha} = 1$. The expectation $E$ of 
the energy for this state is $E=h\cos\theta$, and its variance 
is $h^{2}\sin^{2}\theta$. \par 

With this in mind we now examine the situation where the spin 
one-half particle is immersed in a heat bath. The conventional 
density matrix $r^{\alpha}_{\beta}(T)$ for the system in this 
case can be expressed as a superposition of projection operators 
onto energy eigenstates, with Boltzmann weights: 
\begin{equation} 
r^{\alpha}_{\beta}\ =\ \frac{-e^{-\beta E_{0}}{\bar P}^{\alpha}
P_{\beta} + e^{-\beta E_{1}}P^{\alpha}{\bar P}_{\beta}}
{e^{-\beta E_{0}} + e^{-\beta E_{1}}}\ . \label{eq:qsm} 
\end{equation} 
This can also be written $r^{\alpha}_{\beta} = Z^{-1}
\exp(-\beta H^{\alpha}_{\beta})$. The fact that 
$r^{\alpha}_{\beta}$ has trace unity follows from the 
normalisation condition. For the expected energy 
$H^{\alpha}_{\beta}r^{\beta}_{\alpha}$ we then obtain 
$E=-h\tanh(\beta h)$. \par 

Next we apply the canonical ${\sl\Gamma}$-distribution to this 
problem. For this ensemble the probability distribution of the 
energy in the spin one-half case is 
\begin{equation} 
p(E)dE\ =\ -2\pi Z(\beta)^{-1}\sin\theta 
\exp(-\beta h\cos\theta) d\theta\ , 
\end{equation} 
where $E=h\cos\theta$. The normalisation condition on the density 
function $p(E)$ implies $Z(\beta)=4\pi(\beta h)^{-1}\sinh(\beta h)$, 
from which the expected energy 
$E=-\partial\ln Z/\partial\beta$ can be deduced. The result is 
$E = kT - \mu B \coth\left( \mu B/kT\right)$. Here we have 
reinstated Boltzmann's constant $k$, the particle's magnetic moment 
$\mu$, and the external magnetic field strength $B$, with $h=\mu B$. 
The behaviour of the energy for the cases considered above is 
sketched in Fig. 1. It is clear that the increase in magnetisation, 
when the temperature decreases, is slower in the case of the 
canonical ${\sl\Gamma}$-distribution than for the conventional 
ensemble. This is because the latitudinal circles 
closer to the equator have larger weights than those closer to the 
poles. In Fig. 2 we plot the relationship between the specific 
heat and the temperature. \par 

We note that in the case of the canonical ${\sl\Gamma}$-ensemble 
the heat capacity for this system is nonvanishing at zero 
temperature. For many bulk substances, on the other hand, the heat 
capacity vanishes as zero temperature is approached. Therefore, it 
would be interesting to enquire if a single electron has a different 
behaviour. To that end, we note that in the case of a general two 
state system, with energy eigenvalues $E_{0}$ and $E_{1}$, the 
density matrix of the canonical ${\sl\Gamma}$-distribution, in the 
basis of energy eigenstates, is 
\begin{eqnarray} 
\rho^{\alpha}_{\beta}\ &=&\ -\left[ \frac{1}{\beta(E_{0}-E_{1})} 
+ \frac{1}{1-e^{-\beta(E_{1}-E_{0})}}\right]{\bar P}^{\alpha}
P_{\beta} \nonumber \\ 
& &\ + \left[ \frac{1}{\beta(E_{1}-E_{0})} + \frac{1}
{1-e^{-\beta(E_{0}-E_{1})}}\right]P^{\alpha}{\bar P}_{\beta} . 
\end{eqnarray} 
 
This expression can be contrasted with the conventional result for 
such a system, given in formula (\ref{eq:qsm}). In fact it is not 
difficult to obtain analogous expressions in higher dimensions, 
some of which may characterise other systems that can be 
implemented in experiments to determine whether the 
${\sl\Gamma}$-distribution introduced here provides a better 
account of phenomena in certain situations. \par 

\section{Discussion} 

The foregoing constructions are based on the quantum microcanonical 
postulate, which for an isolated system in equilibrium implies a 
uniform distribution over any energy surface of the quantum phase 
space. We therefore enquire whether this postulate can be derived 
from the basic principles of quantum mechanics. This would follow, 
e.g., if quantum dynamics were ergodic on the energy surfaces 
${\cal E}_{E}$. In the example of the spin one-half particle, 
ergodicity is indeed guaranteed by the periodicity of the 
Schr\"odinger evolution and the dimensionality of the energy surface. 
\par 

In higher dimensions, the generic energy surface ${\cal E}_{E}$ is 
parameterised by $(n-1)$ angular parameters and $n$ phase variables. 
For each fixed set of angular variables we obtain an $n$-torus 
$T^{n}$ in ${\cal E}_{E}$, and by varying the angular parameters the 
energy surface is foliated by an $(n-1)$-dimensional family of 
such tori. If we assume that the Hamiltonian is 
nondegenerate, and that the ratios $E_{i}/E_{j}$ ($i\neq j$) of the 
energy eigenvalues are irrational, then the Schr\"odinger evolution 
is nonperiodic on each $T^{n}$. As a result, ergodicity is 
guaranteed on the toroidal energy subsurfaces. On the 
other hand, unitary evolution does not change the angular parameters 
on ${\cal E}_{E}$. It follows that the energy surfaces are not fully 
ergodic with respect to the Schr\"odinger equation, and that we 
cannot deduce the microcanonical postulate directly from the 
principles of quantum mechanics. It would be interesting to see if 
this postulate could be derived as a consequence of a suitable 
generalisation of ordinary quantum mechanics, e.g., nonlinear 
\cite{kibble,gibbons,ashtekar,weinberg} or stochastic 
\cite{percival} quantum dynamics. \par 

To summarise, we have formulated quantum mechanical analogues 
of the microcanonical and canonical ensembles by consideration 
of the geometry of the quantum phase space. In particular, a key 
distinction in our proposal is that phase space weightings are 
fully taken into account in the construction of equilibrium 
ensembles, whereas in the conventional approach such weights 
arise only in the event of energy eigenvalue degeneracies. We are 
thus able to give a coherent meaning to the quantum microcanonical 
ensemble, and derive an expression for the corresponding density 
matrix, a construction that is missing in the conventional theory. 
It may be that for bulk substances there is a further component 
to the dynamics, perhaps due to internal interactions at low 
temperature, that forces the quantum phase space to contract down 
to the lattice of energy eigenstates, with appropriate 
multiplicities. In that case we recover the conventional theory. 
It would be interesting to determine, by suitable measurements on 
quantum mechanical systems, whether the ensembles introduced here 
are indeed appropriate. \par 

DCB is grateful to PPARC for financial support. \par 

$*$ Electronic address: d.brody@damtp.cam.ac.uk \par 
$\dagger$ Electronic address: lane@ml.com\par

\end{document}